\documentclass[%
 aip,
 jap,
 amsmath,amssymb,
 reprint,%
 floatfix,
longbibliography
]{revtex4-1}

\draft 

\usepackage{color}
\usepackage[english]{babel}
\usepackage{siunitx}
\usepackage{amsmath}
\usepackage{mathtools}
\newcommand*{\red}{\textcolor{black}}
\newcommand*{\2}{\textcolor{black}}
\newcommand*{\rc}{\textcolor{black}}
\usepackage{epsfig}
\usepackage[title]{appendix}
\usepackage{upgreek}

\usepackage{xcolor}
\usepackage{braket}

\definecolor{roed}{RGB}{0,100,180}
\usepackage[normalem]{ulem}
\usepackage[colorlinks=true,allcolors=roed]{hyperref}		

\begin{document}


\title{\rc{Data capacity scaling of a distributed Rydberg atomic receiver array}} 


\author{J.~Susanne~Otto}
\email[]{susanne.otto@postgrad.otago.ac.nz}
\author{Marisol~K.~Hunter}\noaffiliation
\author{Niels~Kj{\ae}rgaard}\noaffiliation
\author{Amita~B.~Deb}\
\email[]{amita.deb@otago.ac.nz}
\noaffiliation
\affiliation{Department of Physics, QSO-Centre for Quantum Science, and Dodd-Walls Centre, University of Otago, Dunedin, New Zealand}


\date{\today}

\begin{abstract}
The data transfer capacity of a communication channel is limited by the Shannon-Hartley theorem and scales as $\text{log}_2(1 + \text{SNR})$ for a single channel with the power signal-to-noise ratio (SNR). We implement an array of atom-optical \rc{receivers} in a single-input-multi-output (SIMO) configuration by using spatially distributed probe light beams. The data capacity of the distributed receiver configuration is observed to scale as \rc{$\text{log}_2(1 + N\times\text{SNR})$} for an array consisting of $N$ receivers. Our result is independent on the modulation frequency, and we show that such enhancement of the bandwidth cannot be obtained by a single receiver with a similar level of combined optical power. We investigate both theoretically and experimentally the origins of the single channel capacity limit for our implementation. 
\end{abstract}

\pacs{}

\maketitle 

\section{Introduction}
In recent years there has been a growing interest in atom-based techniques for the detection of microwave (MW) electric fields \cite{Sedlacek_2012,Fan_2015}, that allow for calibration free SI-traceable measurements \cite{Holloway_2017}\rc{ achieving ultra-high sensitivity\cite{Jing_2020}}. Rydberg atoms have been identified as particularly suitable for such measurements of radio-frequency (RF) electric fields due to their high polarizabilities and large microwave (MW) transition dipole moments \cite{Gallagher_1988}. \2{Advances in \mbox{Rydberg} based field measurements have been accompanied by applications in atom-based communication technology.} The fundamental working principles of analog and digital communication, where a baseband signal is modulated onto an electromagnetic MW carrier wave, have recently been demonstrated in a range of Rydberg-based systems. Examples include amplitude modulation (AM) \cite{Deb_2018,Meyer_2018b,Song_2018,Song_2019,Holloway_2019b}, frequency modulation (FM) \cite{Holloway_2019c,Anderson_2020} and phase detection \cite{Simons_2019,Holloway2019c}, as well as multiple bands \cite{Holloway_2019c}, multiple channels \cite{Zou_2020}, and multiple species \cite{Holloway_2019b}. Common to these methods is the broad carrier frequency range from tens of MHz to \red{several THz} that can be covered by the numerous Rydberg states of a single atomic species.%

Rydberg receivers generally rely on the phenomenon of electromagnetically induced transparency (EIT) \cite{Petrosyan_2011} in a three-level system, where a coupling laser field renders an otherwise opaque atomic medium transparent to a probe laser field due to quantum interference. \2{If the three-level atomic system is coupled to a fourth level via an RF transition, the Autler-Townes (AT) effect \cite{PhysRev.100.703} alters the transmission of a light field.} For an RF electric field, which is modulated with a signal, measuring the modulation of the transmitted light allows to directly retrieve the signal. In contrast to conventional receivers which rely on band-specific electronic components, Rydberg-based receivers benefit from a direct and real-time readout of information, and physical reconfiguration is not necessary when the carrier frequency is varied. Also, the received information is encoded in a light field, i.e., a laser beam, suitable for long-distance transport via a fiber-link.
\begin{figure}[t]
\centering
\includegraphics[trim=0.0cm 0.0cm 0cm 0cm, clip=true,width=0.97\columnwidth]{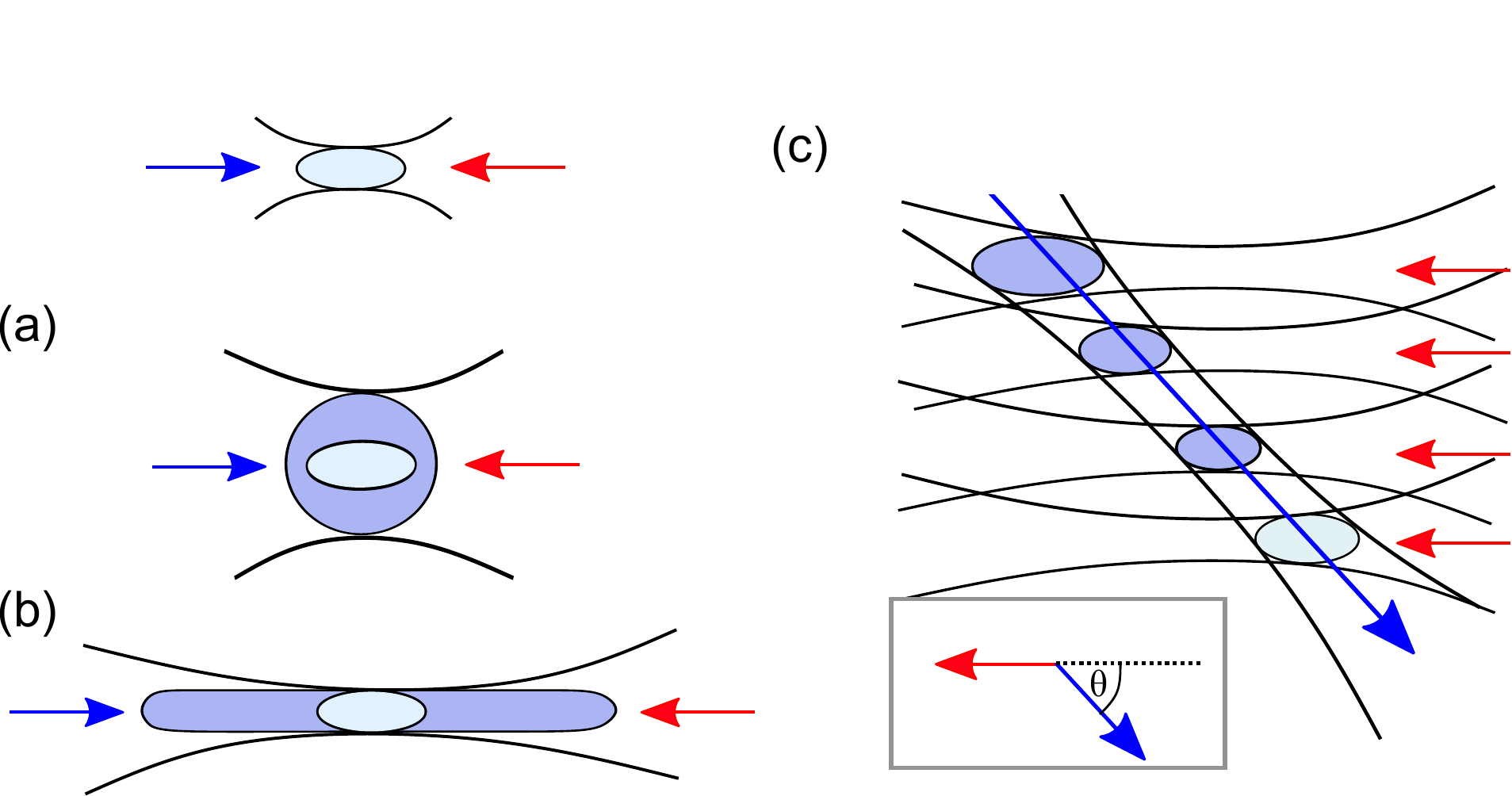}
\caption{\rc{Possible scenarios for increasing the atomic volume of a Rydberg receiver (top). (a) Counter-propagating setup of probe (red) and coupling beam (blue) with larger beam diameters. (b) Counter-propagating beam setup with stretched optical path lengths of probe and coupling beam. (c) Setup with several parallel probe beams and a single coupling beam passing through the probe beams under an angle $\theta >0$. }
\label{fig0}}
\end{figure}

One of the most important figures of merit for a communication system is the channel capacity, which gives the amount of information that can be passed through a communication channel in a period of time without error. For a channel with a bandwidth (BW) its upper limit is defined by the Shannon-Hartley theorem \cite{Shannon_1948} as
\begin{equation}
C = \text{BW}\times \text{log}_2(1+\text{SNR}) \quad (\text{bits/s}).
\label{SHT}
\end{equation}
The crucial parameter for a communication system is therefore the power signal-to-noise ratio (SNR) at a given bandwidth, which has to be maximised in order to attain the maximal channel capacity. \rc{A common method for increasing the data transmission capability is to establish multiple channels using, e.g., multiple transmitter and receiver antennas. A particularly successful technique for fulfilling demands of high data throughput rates and overall capacity are multi-input-multi-output (MIMO) systems\cite{Barry_2004}, and are widely used in telecommunication. MIMO devices employ arrays of receivers which are responsive to multiple simultaneous and independent data streams from multiple transmitter antennas, exploiting multipath propagation. A building block in the context of MIMO technology is a single-input-multi-output (SIMO) arrangement, where the receiver diversity is increased, but the transmitter kept to a single antenna. This arrangement is formally equivalent to a single receiver with an increased SNR, and therefore a higher data capacity is achieved. }For $N$ receiver antennas the data capacity increases with \rc{\cite{Gunasekaran2011}
 \begin{equation}
C_A(N) = \text{BW}\times \text{log}_2(1+N\times\text{SNR}) .
\label{SHT2}
\end{equation}}

In this paper we investigate the scalability of the data capacity of atomic radio receivers in a regime where the receivers are independent of one another and are nearly identical. This, as mentioned above, realises a stepping stone towards MIMO system, where independent and spatially isolated receivers are a necessity. 

As opposed to conventional antennas, Rydberg receivers can work in the electrically small regime \cite{Meyer_2018}. Therefore it is possible, in principle, to improve the data capacity by increasing the atomic density within the receiver volume. In a vapour cell environment, this usually means that one needs to increase the vapour pressure within the cell by heating it which introduces significantly larger Doppler width and collisional broadening of the transition. \rc{Alternatively, the receiver volume can be expanded. This can, for example, be done by using larger beam diameters [see Fig.~\ref{fig0}(a)], but comes with the drawback of requiring higher coupling laser powers to maintain high coupling Rabi frequencies. Another option is to extend the optical path length of probe and coupling beam, see Fig.~\ref{fig0}(b). While this can initially increase the SNR, it also causes the background optical depth, which does not contribute to the signal, to rise, resulting in a drop of SNR. This can be counteracted by increasing the probe beam power to begin with, but the SNR will always \rc{reach a limit} beyond a certain probe beam power for a given coupling laser power. The data capacity for the aforementioned configurations will vary greatly with the geometry of the setup, as the expanded atomic volumes spatially adjoin the original volume and are not necessarily independent of each other, and therefore do not constitute independent receivers. }

\rc{Here we report on an experimental implementation of an array of independent Rydberg receivers that are nearly identical, and therefore accomplish a step towards a MIMO atomic receiver system. This is realised by introducing a small angle between the coupling and probe laser beams as schematically shown in Fig.~\ref{fig0}(c) and without the demand for a higher coupling laser power. We show that this SIMO system follows the well-described data capacity scaling with the number of independent receiver volumes, confirming their independence. Specifically, we employ up to four distributed receiver volumes within a single vapour cell and we observe a log$_2(1+N\times\text{SNR})$ scaling of the data capacity, where $N$ is the number of spatially distributed receivers.}


 \noindent
\begin{figure}[t]
\centering
\includegraphics[width=0.96\columnwidth]{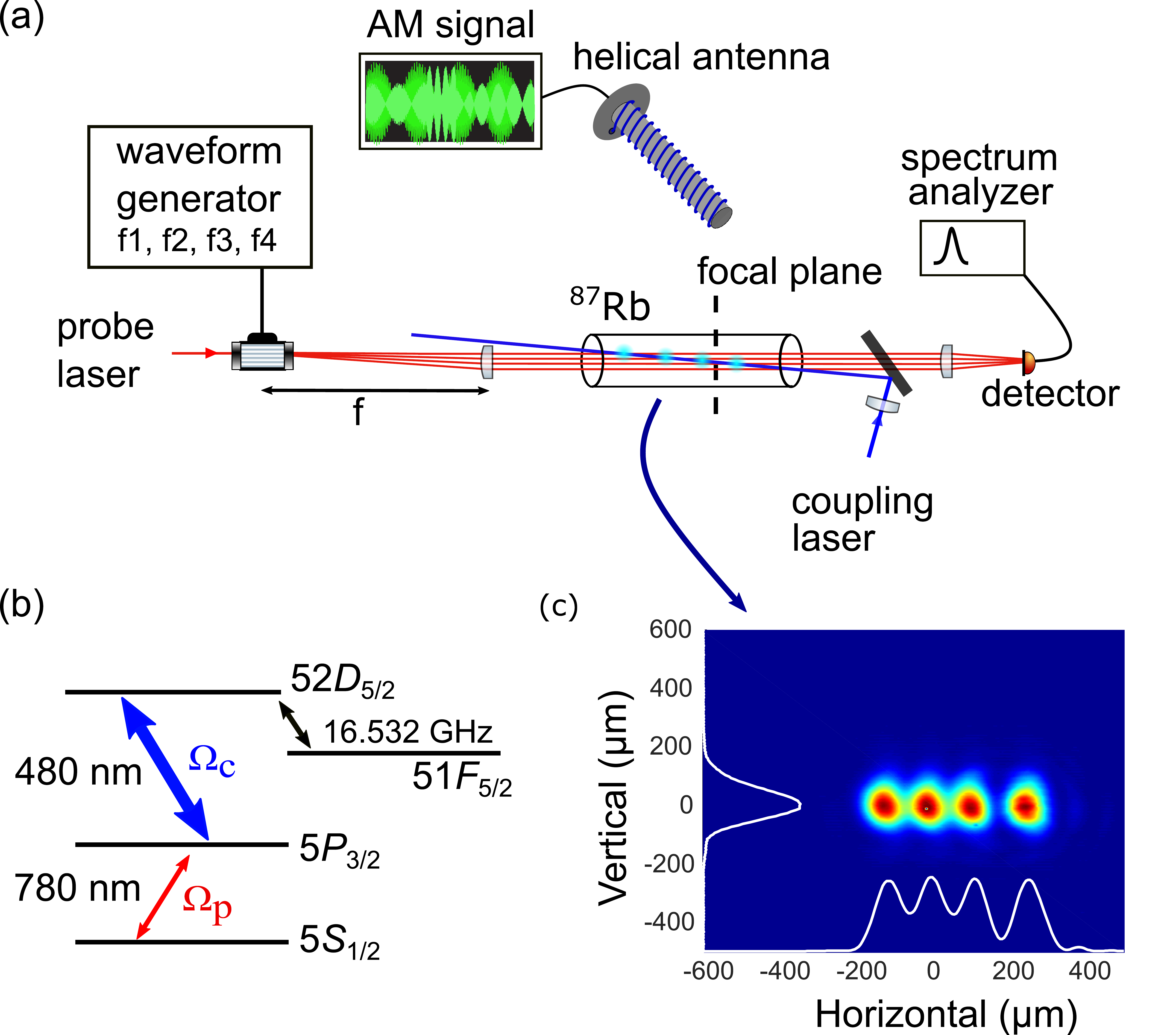}
\caption{(a) Experimental setup used in this work and described in the text. (b) Four-level energy diagram for our atomic $^{87}$Rb system. $\Omega_p$ and $\Omega_c$ denote the probe and coupling Rabi frequencies. (c) Beam profiles of four probe beams in the focal plane with frequency differences $f_1-f_2$ = $f_2-f_3$ = \SI{3}{\MHz} and $f_4-f_3$ = \SI{4}{\MHz}.
\label{setup}}
\end{figure}

\section{Experimental Setup}
The heart of our experimental setup, see Fig.~\ref{setup}(a), is a \SI{75}{\mm}-long,  \SI{25}{\mm}-diameter cylindrical vapour cell containing rubidium (Rb) atoms. The Rb atoms serve as receiver medium for an AM microwave signal emitted by a helical antenna. The AM microwave electric field is optically detected in a ladder-type Rydberg-EIT scheme, where a strong coupling field generates a transparency window on resonance in presence of a weak probe field, see Fig.~\ref{setup}(b). This allows to transduce AM information in the RF domain to optical information encoded as a change of the probe transmission.

In our setup the coupling laser has a wavelength of \SI{480}{nm} and a power of \SI{22}{\mW}, and is focused to a waist of \SI{60}{\micro m} in the vapour cell. This corresponds to a Rayleigh length of $\sim\SI{25}{mm}$ and a Rabi frequency of 2$\pi\times$\SI{8}{\MHz} for the transition 5$P_{3/2} \leftrightarrow$ 52$D_{5/2}$. A probe light field at \SI{780}{\nm} passes through an acousto-optic modulator (AOM), which is driven by a multitone frequency source. Up to four diffracted beams at frequencies $f_1, f_2, f_3$ and $f_4$ can be generated simultaneously. The diffracted beams impinge on a lens which generates parallel beams with $\sim$\SI{70}{\micro m} radii in its focal plane, see Fig.~\ref{setup}(c). The probe and coupling beams are counter-propagating under an angle of $\sim$\SI{2}{\degree} to create spatially separated ($>$ \SI{3}{\mm}) \rc{and independent} overlap areas between the coupling and probe beams. 
The transmission signal of the probe beams is collected by a lens and focused onto a photodetector. For the purpose of retrieving the frequency and amplitude of the modulation signal the photodetector is connected to a spectrum analyser. 
\newline
\indent
For the frequency stabilisation of the probe and coupling lasers, two auxiliary Rb vapour cells are employed. The probe laser is stabilized to a saturated absorption spectroscopy signal of the $^{87}$Rb D$_2$-line. With the AOM in front of the experimental cell, see Fig.~\ref{setup}(a), we obtain four beams with frequencies close to the transition $5S_{1/2}(F=2) \rightarrow$ $5P_{3/2} (F'=3)$. For the stabilisation of the tuneable coupling laser the EIT signal of a $\upmu$-metal-shielded Rb vapour cell is used, and the laser can be stabilized to different Rydberg levels $nD_{5/2}$ and $nD_{3/2}$ with $n = 30$ to $70$.  For the results presented in this paper we use the Rydberg level 52$D_{5/2}$.
\newline
\indent
Our experimental vapour cell is kept at a temperature of \SI{85}{\degree C} yielding a ground state atom density of $\sim10^{12}\SI{}{cm^{-3}}$ for a $^{87}$Rb sample. The two optical fields which are passing through the cell are coupled to the Rydberg level 51$F_{5/2}$ with a MW carrier field at \SI{16.532}{\GHz}.  This field is generated by an analog signal generator that feeds a \SI{10}{dBm} signal into a home-built helical end-fire antenna. In axial mode the antenna radiates microwave fields along the helical axis and the radiation is circularly polarized. The antenna is located \SI{30}{cm} from the vapour cell. With increasing field strength of the MW field, the amplitude of the EIT signal decreases while its width broadens and for MW electric fields $>$\SI{10}{dBm} our transmission signal splits into two AT peaks. The MW carrier field is amplitude modulated with a sinusoidal signal with frequencies between \SI{100}{\kHz} and \SI{1}{\MHz}. This results in a variation of the transmission of the probe fields, which can be directly measured with the combination of photodetector and spectrum analyser and permits to retrieve the initial AM signal. 
\section{Results}
\label{sec3:1}
\subsection{SNR and Bandwidth of a Single Receiver}
\label{sec3a}
\begin{figure}[t]
\centering
\includegraphics[trim=0.0cm 0.0cm 0cm 0cm, clip=true,width=0.90\columnwidth]{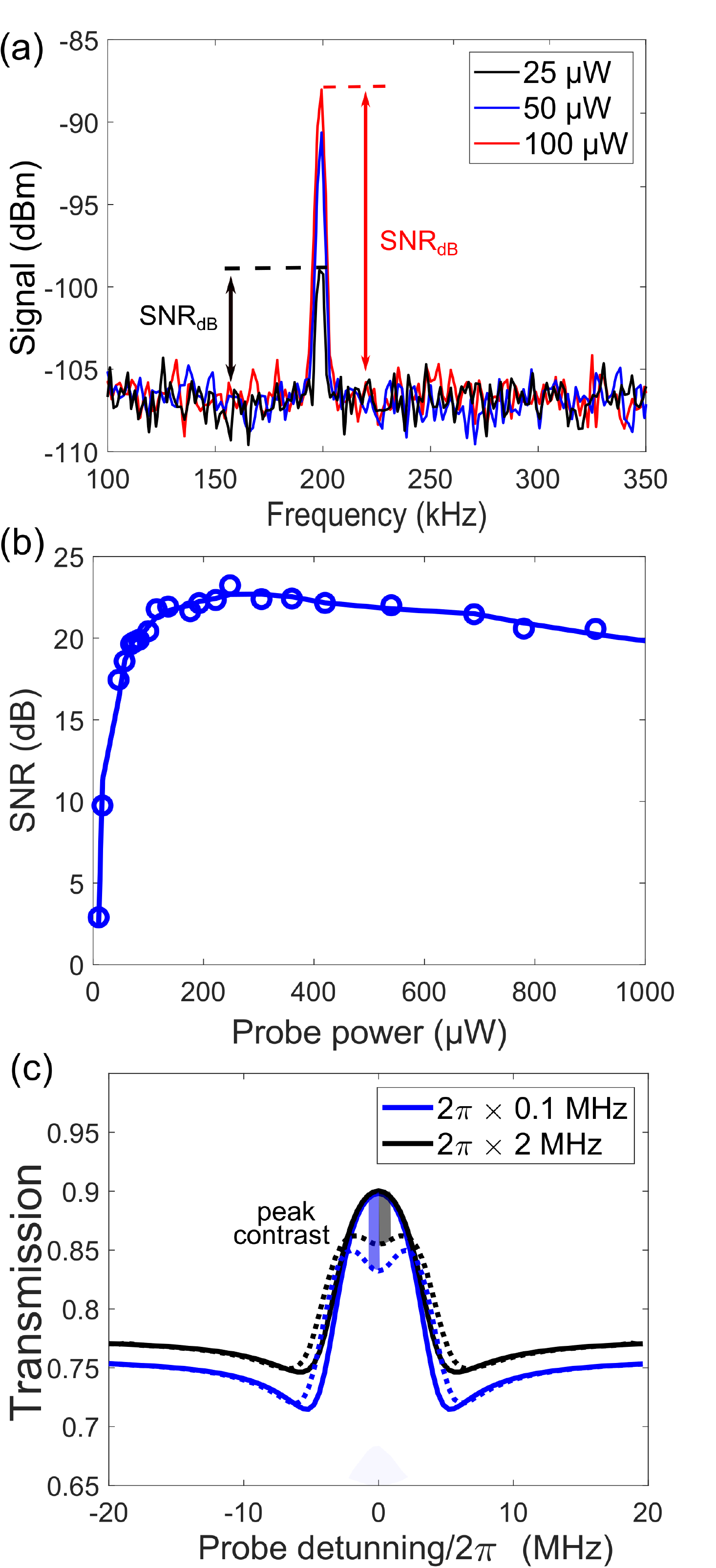}
\caption{(a) Demodulated signal of the probe transmission as a function of frequency measured with the spectrum analyser for three different probe powers. (b) Dependence of the SNR$_\text{dB}$ on the probe power for an AM frequency of \SI{200}{\kHz}. (c) Numerical data for the probe detuning versus probe transmission for two different probe Rabi frequencies (blue and black), in the EIT regime (solid lines) and AT regime (dotted lines) for a resonant RF field ($\Omega_\text{RF} = 2\pi \times \SI{5}{\MHz}$).
\label{spectrumscan}}
\end{figure}
Figure~\ref{spectrumscan}(a) shows the demodulated probe signal for a carrier field amplitude modulated at \SI{200}{kHz}, measured with the spectrum analyser for a resolution bandwidth of \SI{3}{\kHz}. The transmitted probe field is detected by the photodetector, and the spectrum analyser yields the power spectral density for frequencies between 100 and \SI{350}{kHz}. The traces have distinct maxima at the modulation frequency, which are standing out against the noise floor observed with the spectrum analyser. We define the SNR$_\text{dB}$ as the difference of the power spectral density at the modulation frequency to the noise floor on the spectrum analyser at the respective frequency in absence of the modulation. As can be seen from Fig.~\ref{spectrumscan}(a), the signal height and \rc{along} with it the SNR$_\text{dB}$  grow with increasing probe power in the presented scenario (black to red). 
\begin{figure*}[t]
\centering
\includegraphics[width=1.97\columnwidth]{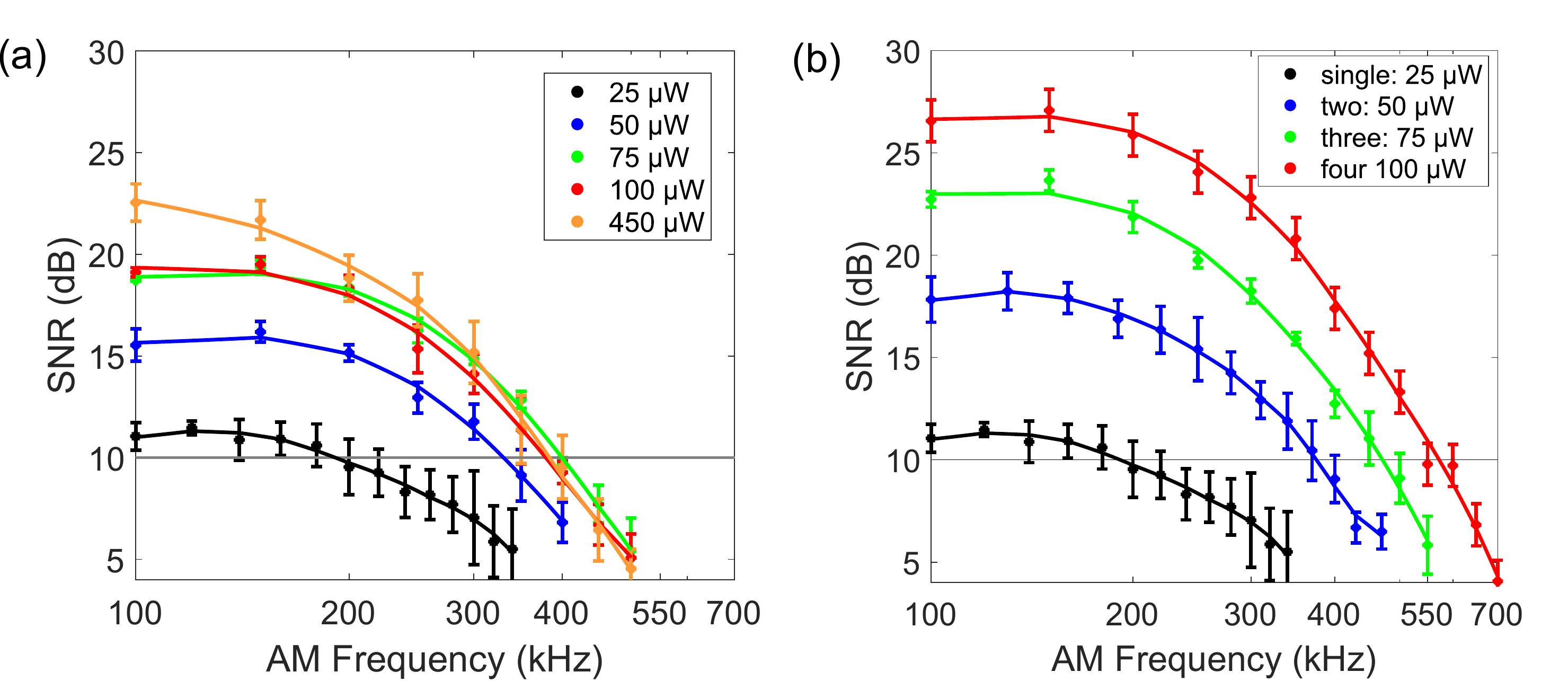}
\caption{\label{sBeam}(a) SNR$_\text{dB}$  as a function of AM frequency of the carrier MW field for different power levels of a single probe beam. The plot shows average values of four probe beams with frequency offsets to the resonance of \SI{-4.5}{\MHz}, \SI{-1.5}{\MHz}, \SI{1.5}{\MHz} and \SI{5.5}{\MHz}. (b) SNR$_\text{dB}$  as a function of AM frequency for combinations of up to four probe beams of \SI{25}{\micro W} with frequency offsets to the resonance frequency of $f_1$ = \SI{-4.5}{\MHz}, $f_2$ =\SI{-1.5}{\MHz}, $f_3$ =\SI{1.5}{\MHz} and $f_4$ =\SI{5.5}{\MHz}. Measurements were taken for all combinations of $f_1, f_2, f_3, f_4$ and average values and corresponding standard deviations are presented. \rc{The curves, obtained from a Savitzky–Golay filter, are to guide the eyes. }}
\end{figure*}
\noindent
The development in SNR$_\text{dB}$  as a function of probe power is shown in more detail in Fig.~\ref{spectrumscan}(b). \rc{Initially, the SNR$_\text{dB}$ grows with increasing probe power, but reaches a maximum at $\sim\SI{150}{\uW}$, after which it slowly decreases }for even higher probe powers\rc{, if the experimental conditions, i.e. atomic density and coupling power are kept identical.} The growth of the SNR$_\text{dB}$  with probe power is related to an increase in peak contrast, as illustrated in Fig.~\ref{spectrumscan}(c), which is the difference between the probe transmissions at the EIT peak in presence and absence of the carrier RF field. In the four level scheme in Fig.~\ref{setup}(b), the applied RF field suppresses the EIT transmission on resonance as shown as dotted lines in Fig.~\ref{spectrumscan}(c) for two different probe Rabi frequencies (black and blue). For strong RF fields the EIT transmission on resonance is suppressed and can reach the baseline, which is given by the absorption of the probe beam without RF and coupling field.

The development of peak contrast with probe power ($\propto \Omega_p^2$) can be qualitatively explained as follows: For larger $\Omega_p$, power broadening leads to an increase of the spectral width of the atomic absorption line, resulting in an increased transmission of the probe beam and a rising baseline transmission. At the same time, for a fixed Rabi frequency of the coupling field, $\Omega_c$, and weak $\Omega_p$ the population of the meta-stable Rydberg state increases with $\Omega_p$, since the population ratio between the ground and Rydberg state is determined by $\Omega_p^2/\Omega_c^2$. If less atoms are in the ground state the transmission of the probe field increases. Overall this results in a larger EIT peak \cite{Wu_2017}, and the observed initial rise of the SNR$_\text{dB}$  in Fig.~\ref{spectrumscan}(b). \rc{As the number of atoms in the receiver volume is kept constant, the EIT peak transmission saturates \cite{Hao_2018} for even larger $\Omega_p$ while the two-level baseline transmission continues to grow, see Fig.~\ref{spectrumscan}(c).  In our setup the noise floor on the spectrum analyser rises as a quadratic function of the probe power for $>\SI{50}{\uW}$, see Fig.~\ref{noisefloor} in the Appendix~\ref{AA}, and limits the achievable SNR$_\text{dB}$ \rc{for higher probe powers. The dominant mode of noise in our experiment is introduced by an AOM which is far above that posed by photon shot noise (more discussed in Appendix~\ref{AA}). This explains the slow fall off of the SNR$_\text{dB}$ for probe powers $>\SI{150}{\uW}.$} } %

In addition to the probe power, the achievable SNR$_\text{dB}$ depends on the modulation frequency of the carrier RF field, as shown in Figure~\ref{sBeam}(a). The fall-off of the SNR$_\text{dB}$  with increasing AM frequency sets the bandwidth of the receiver system. 
We define the BW limit as the cutoff point with SNR$_\text{dB}$  = \SI{10}{dB}. Figure~\ref{sBeam}(a) shows a measurement of the BW for a single probe beam and five different probe powers. For faster modulations the SNR$_\text{dB}$  decreases and reaches its BW limit at \SI{380}{\kHz} for probe powers $\geq \SI{100}{\uW}$.
The slope of the SNR-curves is determined by atomic parameters such as decoherence rates set by Doppler broadening and transit times, which are the same for all SNR-curves in Fig.~\ref{sBeam}(a). Increasing the SNR$_\text{dB}$ initially improves the cut-off frequency, \rc{but as the SNR$_\text{dB}$ reaches its maximum the BW saturates.}

\subsection{Multiple Atomic Receivers}
\label{sec3b}
\begin{figure*}[t]
\centering
\includegraphics[width=1.99\columnwidth]{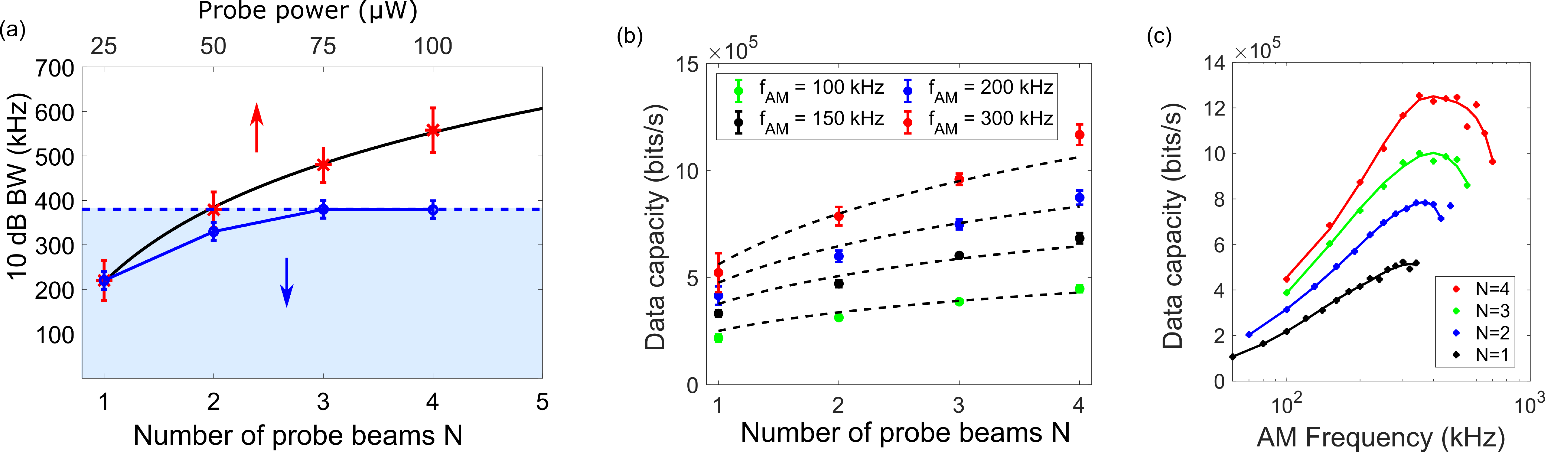}
\caption{(a) \SI{10}{dB} bandwidth limit of a single probe beam as function of the probe power (blue, bottom axis) and for multiple beams (red, top axis), where the probe power is distributed equally among $N$ beams. The BW increases with logarithmically (black line), and surpasses the BW limit of the single beam (shaded area). (b) \rc{Data capacity for four different AM frequencies and up to four probe beams. The dashed lines show the theoretical scaling of Eq.~(\ref{eq22}), where $\text{SNR}_{N=1}$ was calculated from the experimental data for $N=3$. (c) Data capacity versus amplitude modulation frequency for $N$ beams. } \label{inset}}
\end{figure*}

\rc{With the objective to  obtain independent detection volumes within our vapour cell, 
we distribute the power of our single probe beam over multiple probe beams. In combination with a single coupling beam we obtain up to four simultaneous and nearly identical receiver volumes.} We use AOM driving frequencies with $\Delta f = 3-\SI{4}{\MHz}$ between adjacent \rc{probe} beams and observe a spatial separation of $\sim$\SI{100}{\um} in the focal plane, see Fig.~\ref{setup}(c). We establish a single-input-multi-output (SIMO) configuration, as it is used in the context of smart antenna technology for improved wireless communication performance. 
Figure~\ref{sBeam}(b) shows the SNR$_\text{dB}$  versus AM frequency for one, two, three and four probe beams with individual powers of \SI{25}{\uW} at their probe frequencies $f_1$ to $f_4$. 
The SNR$_\text{dB}$  drops towards higher AM frequency, similar to the scenario for a single receiver in Fig.~\ref{sBeam}(a), since the slope of the SNR-curves is determined by the atomic parameters of our experimental setup. Crucially, however, the saturation of the \rc{EIT peak transmission} for a single beam can be avoided. This allows us to exceed the BW limit of the single beam setup, as shown in Fig.~\ref{inset}. While for two beams the BW limit approximately matches the scenario of a single beam with twice the power, a clear improvement in BW appears for $N \geq 3$, where $N$ is the number of beams. \rc{Overall, the maximum values of the SNR$_\text{dB}$ and bandwidth for the single beam setup can be exceeded}. 
\noindent

Assuming that two receiver areas are independent and have identical parameters such as $\Omega_p$, $\Omega_c$, beam frequencies, waists, and atomic densities, the signal amplitude increases by a factor of two. We note that doubling the optical power that contributes to the signal increases the SNR$_\text{dB}$ by a factor of 4 (roughly \SI{6}{dB}), as e.g. found in Fig.~\ref{sBeam}(a) (black to blue line). This is because the optical power $P_\text{op}$ hitting the photodetector translates into a voltage $V \propto P_\text{op}$, and the spectrum analyser measures the corresponding electric power $P_\text{el}$, for which $P_\text{el} \propto V^2$. In Fig.~\ref{inset}(a) we show the increase in BW for up to four beams. \rc{As expected (see Appendix~\ref{AB}) the BW scales logarithmically (black line), and exceeds the maximum  bandwidth of the single receiver (blue shaded area).} 

\rc{The achievable data capacity for our system with $N$ independent receiver volumes at a given amplitude modulation frequency $f_\text{AM}$ transforms from Eq.~(\ref{SHT2}) to }
\rc{
\begin{equation}
\label{eq22}
C_A(N) = f_\text{AM}\times \text{log}_2(1 + N\times\text{SNR}_{N=1} ).
\end{equation}
}
Figure~\ref{inset}(b) presents the data capacity for three different modulation frequencies and up to four beams. The data capacity shows significant enhancement, following the predicted behaviour of Eq.~(\ref{eq22}) (dashed lines). Moreover the data capacity depends on the amplitude modulation frequency, as presented in Fig.~\ref{inset}(c). For low AM frequencies the data capacity has a linear dependence on the modulation frequency. For faster modulations the data capacity reaches a maximum, before decreasing due to the decreasing SNR$_\text{dB}$  caused by a finite atom-switching time. \rc{The maximum of the data capacity moves towards higher modulation frequencies with the number of probe beams, $N$, as the SNR$_\text{dB}$ increases with $N$.}

\rc{The maximum values for bandwidth ($\sim\SI{560}{kHz}$) and data capacity ($\sim\SI{1.3}{MBits/s})$ achieved in this setup are rather low in comparison to other experiments\cite{Meyer_2018b}. This is predominantly caused by the angled geometry of the setup (see Fig.~\ref{fig0}), and is discussed in more detail in Sec.~\ref{Sec:3ca}. Importantly, \rc{for our currentstudy} bandwidth and channel capacity scale in a predicable manner with the number of receiver volumes, as the receiver volumes are independent.} 


\subsection{Characterisation of the Distributed \red{Receivers} Setup }
\label{Sec:3c}

\subsubsection{\rc{Effects of the angled configuration }}
\label{Sec:3ca}
\rc{The coupling laser power for our setup is limited to \SI{22}{mW}. For the purpose of achieving independent receiver volumes without a need for several coupling laser beams, the coupling and probe laser beams are configured to counter-propagate under a small angle, so that one coupling beam is used for all receiver volumes. The angled geometry comes with the drawback of decreasing the SNR for the independent receiver volumes. This is a consequence of an additional Doppler shift\cite{Carvalho2004}, which broadens the linewidth of the EIT signal. In our setup, see Fig.~\ref{fig0}(c), an angle of $\theta \approx \SI{2}{\degree}$ results in a EIT linewidth that is approximately twice as large as for $\theta = \SI{0}{\degree}$. However, while we do see a broadening of the EIT linewidth for $\theta > \SI{0}{\degree}$, we do not see any additional change while using multiple probe beams. Therefore, we realise a setup with independent receivers as different groups of atoms are addressed within the vapour cell at the same instant of time. Another factor to consider is probe signal loss due to scattering of the probe light with atoms outside the sensing area. The length of the sensing area is with $\sim\SI{1}{mm}$ small in comparison to the full path length of the probe beams in the \SI{75}{mm} long vapour cell.\footnote{In configurations with $\theta = \SI{0}{\degree}$ the overlap area of the probe and coupling beam in the vapour cell is larger. However, as the coupling and probe beam are usually focused to a small spot size, the volume contributing the most to the EIT signal is comparatively small due to the shorter Rayleigh lengths.} This problem could be avoided in future realisations by using miniaturised vapour cells.  }

\rc{The scope of this paper is to investigate the scalability of the data capacity for a Rybderg-atom based SIMO system. Therefore the independence of the the receiver volumes is of greater significance than high signal-to-noise ratios. With an average atomic velocity of $\sim \SI{150}{m/s}$, adjacent receiver spacings exceeding $\SI{3}{mm}$, and amplitude modulation frequencies of $>\SI{100}{kHz}$, interactions due to atoms transiting between sensor areas are negligible. Because of the angled geometry the SNRs of the atomic volumes are overall low, but the configuration allows to establish the scaling law for the channel capacity of a SIMO setup.   }


\begin{figure}[b]
\includegraphics[trim=0.0cm 0.0cm 0.0cm 0.0cm, clip=true,width=0.98\columnwidth]{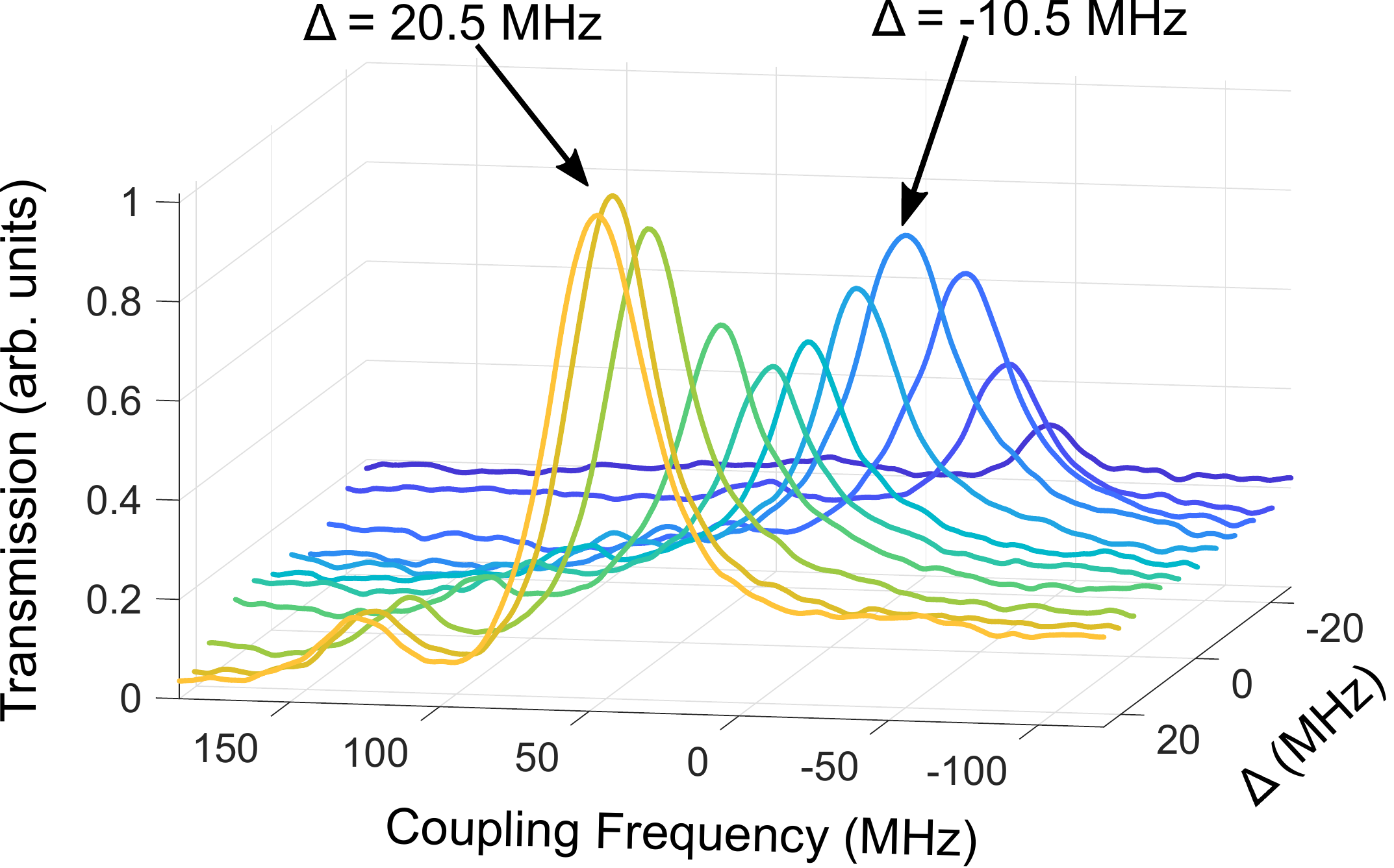}
\centering
\caption{Probe transmission for different AOM driving frequencies $\Delta$ (EIT resonance at $\Delta$ = 0) at a fixed probe power of of \SI{300}{\uW}. The coupling laser frequency is scanned and the frequency axis calibrated using the known splittings between the states $52D_{5/2}$ (large peaks) and $52D_{3/2}$ (smaller peaks on the left side). \label{EIT} }
\end{figure}
\subsubsection{Deviations of EIT peak heights for different receiver volumes}
Unlike in an idealized scenario with identical distributed receivers, in our experimental setup the receiver volumes differ slightly due to the geometry of the setup and since the probe power is distributed via an AOM. Therefore, we show average values of all four probe beam combinations in Figs.~\ref{sBeam} and \ref{inset}. Deviations between the detected SNR$_\text{dB}$ for different probe frequencies are caused by a change in the EIT condition. The main contribution comes from slightly different beam waists in the overlap area, affecting the ratio $\Omega_c/\Omega_p$. To illustrate the change of the EIT transmission for different probe frequencies, we present in Fig.~\ref{EIT} the EIT profiles for a scan of the coupling beam frequency. The AOM-offset of the probe frequency with respect to the two-level resonance frequency is denoted as $\Delta$. For all frequencies we observe two EIT peaks, since the EIT condition can be met for the 52$D_{5/2}$ and 52$D_{3/2}$ Rydberg states. Relevant in this investigation is the larger peak. For a fixed probe power of $\SI{300}{\uW}$ 
the height of the EIT peak changes with $\Delta$, since the position of the probe beam in the focal plane moves as $\sim \Delta \times$\SI{37}{\um/MHz}. Considering the angle between the counter-propagating probe and coupling beam, and a scan of $\Delta = \SI{40}{MHz}$ the overlap areas are separated by up to $\SI{40}{mm}$, which is approximately twice as large as the Rayleigh lengths of the probe and coupling beam. The appearance of two maxima at $\Delta = -\SI{10.5}{\MHz}$ and $\Delta = \SI{20.5}{\MHz}$ suggests that the focal plane of probe and coupling beam do not coincide. For the probe frequencies $f_1$ to $f_4$ of Fig.~\ref{sBeam}(b) the EIT peak height differs by \SI{10}{\percent} for a fixed probe power, which translates to deviations in the SNR$_\text{dB}$. This technical limitation can be mitigated, for example by adjusting the probe power of individual beams. 

\subsubsection{SNR$_\text{dB}$ for different probe beam frequencies }
\begin{figure}[t]
\centering
\includegraphics[trim=0.1cm 0.0cm 0cm 0cm, clip=true,width=0.95\columnwidth]{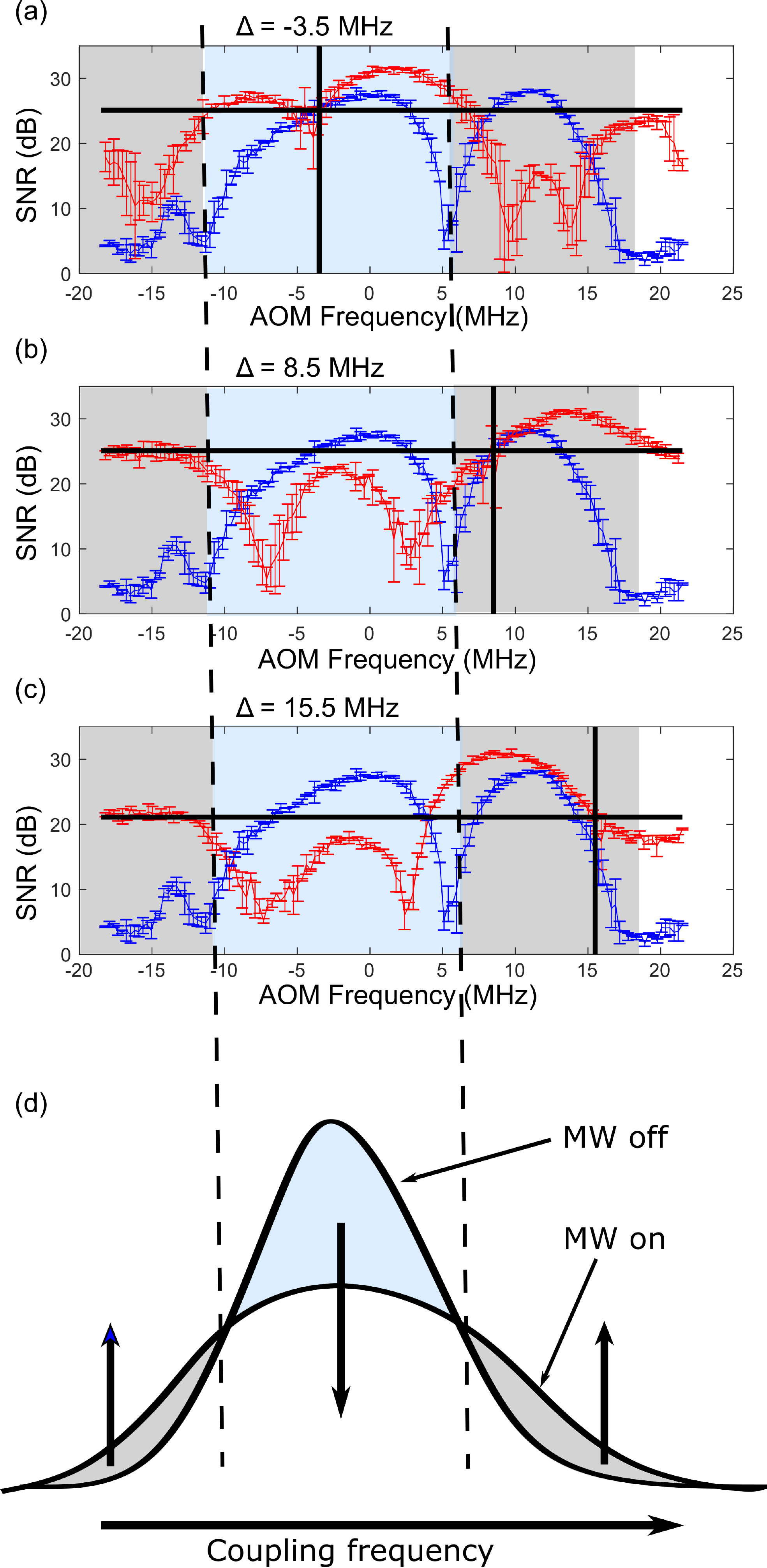}
\caption{\red{SNR$_\text{dB}$ as function of the AOM frequency for a single beam (blue) and an additional beam (red) with frequency detuning (a) -\SI{3.5}{\MHz}, (b) \SI{8.5}{\MHz}, and (c) \SI{15.5}{\MHz} - represented by the vertical black lines. The SNR$_\text{dB}$ of the beam at frequencies (a)-(c) is represented as horizontal black line. A schematic representation of the EIT spectrum with and without MW field for a scan of the coupling laser frequency is shown in (d).  \label{sweep}}}
\end{figure}

With the objective to find \red{optimised} probe frequencies for our system, the coupling laser was locked to the $5P_{3/2} \leftrightarrow 52D_{5/2}$ transition and the probe frequency swept over a \SI{40}{\MHz} range using the AOM. In difference to the measurements in Fig.~\ref{EIT}, we allow for changes in the probe power with AOM driving frequency, which is caused by the dependence of the AOM efficiency on the driving frequency. The SNR$_\text{dB}$  \red{was determined as a function of the AOM frequency (probe frequency) for an amplitude modulated MW field at \SI{100}{kHz}. The results are presented in Fig.~\ref{sweep}(a)-(c) as blue data points}. In our frequency scan, we observe two minima ($\Delta = \SI{-12}{\MHz}$ and $\Delta = \SI{5}{\MHz}$), which can be associated with points for which the EIT transmission does not change in presence of the MW field. High SNR$_\text{dB}$  occurs for the largest change in probe transmission due to the applied MW field. 
For a symmetric AT profile, as depicted in Fig.~\ref{sweep}(d), we expect a symmetric scenario in Fig.~\ref{sweep}(a)-(c) with three peaks with a vertical line symmetry at the resonance frequency $\Delta = 0$, and e.g. observed in \cite{Zou_2020}. A number of factors can lead to the observed asymmetry in Fig.~\ref{sweep}(a)-(c). First, with our MW electric field we intend to couple the Rydberg states 52$D_{5/2} \leftrightarrow 51F_{7/2}$ which has transition dipole moments up to $\SI{2260}{a_0e}$. However, a second Rydberg state, $51F_{5/2}$, is located only $\SI{1.2}{\MHz}$ away and can shift the AT-profile. 
\noindent
\noindent
Secondly, the probe frequency is scanned using an AOM with maximised diffraction efficiency at the resonance $\Delta = 0$. Lastly, the EIT peak height decreases rapidly for frequencies $\Delta  < -\SI{10}{\MHz}$, as shown in Fig.~\ref{EIT}, due to weaker overlap of coupling and probe beam. This explains the difference in height of the left and right peak in Fig.~\ref{sweep}(a)-(c).
\subsubsection{\rc{Effect of an additional probe beam }}
\label{Sec:3c2}
In order to characterise the performance with multiple probe beams, we scanned the frequency of the first probe beam in presence of a second probe beam at three fixed frequency detunings -\SI{3.5}{MHz}, \SI{8.5}{MHz} and \SI{15.5}{MHz}. We measured the SNR$_\text{dB}$ of the two beams by combining the two probe signals on the photodetector. The obtained values of SNR$_\text{dB}$ are presented in red in Fig.~\ref{sweep}(a)-(c). One could naively expect the SNR$_\text{dB}$  to remain at least at the value given by the fixed-frequency beam, depicted in Fig.~\ref{sweep}(a)-(c) as horizontal black lines. Instead, for two probe beams we \red{predominantly} observe a drop in the SNR$_\text{dB}$ if the two probe frequencies are in differently shaded frequency areas (grey and blue), and a rise for two frequencies in identically shaded areas. A descriptive explanation for this phenomenon is given in Fig.~\ref{sweep}(d). While the grey and blue areas both show a significant change of the EIT signal in presence of a MW electric field, for grey areas the probe transmission increases in height, whereas the signal drops in the blue area. By detecting both beams on the same photodetector the transmission signals from two beams can be effectively ``out of phase". If both components are equally strong, a zero occurs in the SNR$_\text{dB}$. Otherwise, the stronger signal dominates the scenario. Hence, for our experimental setup it is crucial to choose frequencies within identically colored areas. All beams can then be detected with a simple detection scheme, involving only a single photodetector. Alternatively, the probe transmission signal can be detected with individual photodetectors to allow for less sensitivity in the choice of probe frequencies. For the SNR$_\text{dB}$  measurements of up to four distributed beams, we used a separation of $3-\SI{4}{MHz}$ for adjacent probe beams centered around $\Delta = 0$. With this choice the individual beams have similar SNR$_\text{dB}$ levels, as well as spatially separated overlap areas. 

We note that the frequency deviations of the probe beams in or setup could be of potential benefit for a MIMO configuration. For the generation of multiple probe beams with identical frequencies electromechanically driven mirrors or liquid crystal deflectors could be implemented.

\section{Discussion}
The fastest switching time in an atom-based receiver is given by the contrast of optical transmission and the detection noise. Using purely classical light sources, the noise floor is ultimately limited by the photon shot noise, while the transmission contrast is determined by the lifetime of a single atom in a dark state, the number of contributing atoms in the EIT process, and the input probe beam power. The atom-switching time in our realisation is primarily limited by \rc{the available coupling laser power. This can be expressed in terms of the EIT pumping rate $\Omega_\text{EIT} = \Omega_c^2/2\Gamma$ \cite{Meyer_2018b}, which describes the time atoms need to re-establish the EIT dark state when the MW field is turned off.}

For a given realisation with a particular atom-switching time, one expects that increasing the probe power or the number of the participating atoms within the receiver volume would lead to a larger contrast in optical transmission, and therefore increase the SNR. We observe that the SNR \rc{reaches a maximum} beyond a certain probe power and \rc{for a given atomic volume}, but by spatially distributing the probe power to address different atoms, one can increase the SNR, which ultimately leads to an increase of the data capacity. Our results demonstrate a scaling of the data capacity with \rc{$C_\text{SIMO} = \text{BW}\times \text{log}_2(1+N\times\text{SNR})$} for $N$ probe beams. \rc{For a single receiver with a probe power that equals the sum of the distributed probe beams power broadening of the intermediate state occurs. Furthermore, a higher fraction of atoms are excited to the Rydberg state, which potentially results in additional Rydberg dephasing. These effects are avoided by distributing the probe power among several receivers. Fundamentally, the capacity advantage of the distributed receiver stems from a larger sensing volume, i.e a larger number of atoms contributing to the signal. }\rc{While we exceed the BW of our single receiver and our data follows the predicted scaling of a SIMO system, the overall achieved data capacity remains low. This originates from low SNR levels of the individual receiver volumes, caused by the angled geometry of our setup, as discussed in Sec.~\ref{Sec:3ca}.} 

Our results particularly highlight the challenge of scaling up the data capacity of Rydberg atomic receivers by classical means, such as scaling up the receiver volume. To put this in perspective, in order to increase the data capacity by an order of magnitude, one needs to implement more than thirty distributed receivers with the receiver volume larger by the same ratio. Assuming a probe volume of a single receiver to be $\sim$ \SI{0.3}{mm^3}, this would require a distributed receiver with a volume of $\sim$ \SI{1}{cm^3}. This shows the importance of maximising single-channel data capacity by reducing the detection noise floor, \rc{Doppler broadening} or by employing collective effects in atomic excitation/de-excitation \cite{Dicke1954,Kwong2014}. Such collective effects have so far remained elusive in warm atomic vapour, but have been demonstrated in cold atomic gases \cite{M_hl_2020}.

\section{Conclusion and Outlook}
We have considered a simple setup employing spatially distributed atomic receivers which allows to surpass the SNR limit of an individual atom-based receiver, resulting in an improved channel capacity. The concept of receiver arrays has proven beneficial in wireless communication systems for high-speed transmission and increased capacity. For example for a SIMO system, as presented in this work, an increase in channel capacity to \rc{$C_\text{SIMO} = \text{BW}\times\text{log}_2(1+N\times\text{SNR})$} \cite{Ghayoula_2014} for $N$ receiving antennas has been \red{derived}. We experimentally confirmed the same scaling for the channel capacity of our atomic-receiver system with $N$ probe beams (receiver volumes). Moreover, we observed a significant increase in SNR$_\text{dB}$, and BW -- the latter scaling logarithmically for our Rydberg atom-based RF-to-optical receiver. We showed that such enhancement of the BW cannot be obtained by a single receiver with an \rc{identical beam geometry and a} similar level of combined optical power due to the saturation of the \rc{EIT peak transmission}. Our approach benefits from the little additional resource overhead needed to add multiple probe beams, if generated by the first order of an AOM and a multitone frequency source. While the results shown in this work were carried out with $N \leq 4$ probe beams, the number of beams can easily be expanded \cite{Roberts:14}. 

\rc{In the future we plan to extend the system with additional transmitter RF antennas in a configuration where the receivers can distinguish between transmitters. This will allow us to realise a  multi-input-multi-output (MIMO) system, promising further enhancement in channel capacity. For an additional transmitter antenna, we expect an improvement in channel capacity with \rc{$2 \times \text{log}_{2}(1 +N\times\text{SNR})$}. This however requires that the spatial separation between the distributed atomic receivers to be in the order of the RF wavelength \cite{Barry_2004}. Furthermore, we might be able to boost the coupling beam power for our setup e.g. using power buildup in a resonant cavity. }

\begin{acknowledgments}
We acknowledge funding from the Marsden Fund of New Zealand (Contract No. UOO1729) and MBIE (Contract No. UOOX1915).
\end{acknowledgments}

\section*{DATA AVAILABILITY}
The data that support the findings of this study are available from the corresponding author upon reasonable request.

\appendix
\begin{figure}[t]
\centering
\includegraphics[trim=0cm 0cm 0cm 0cm, clip=true,width=0.95\columnwidth]{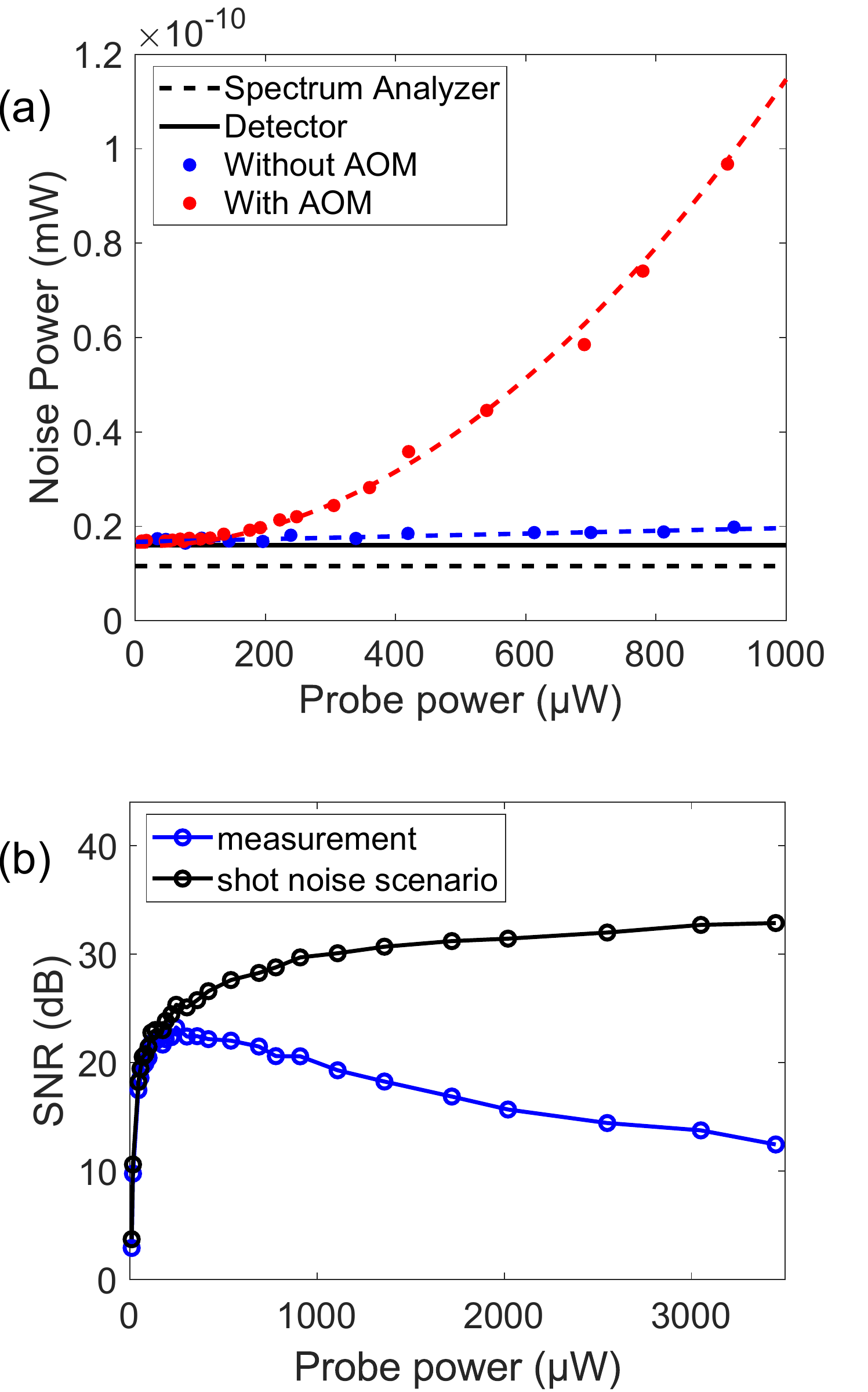}
\caption{(a) Measurement of the noise power on the spectrum analyser over a frequency band of 50 to \SI{350}{kHz} for a resolution bandwidth of \SI{3}{kHz} and different probe powers. The data points show average values over the frequency range for the first order of the AOM (red) and for a measurement without AOM (blue). The noise floor of the photodetector (black line) lies $\sim \SI{1.5}{dB}$ above the noise floor of the spectrum analyser (dashed line). (b) Development of SNR$_\text{dB}$  as function of probe power as shown for our setup in Fig.~\ref{spectrumscan}. If only shot noise is present (black), the SNR$_\text{dB}$ \rc{reaches a maximum of $\sim \SI{32}{dB}$.}   \label{noisefloor}}
\end{figure}
\section{Noise assessment of the AOM}
\label{AA}
The key component for the generation of multiple probe beams in our experimental setup is a continuously running AOM, see Fig.~\ref{setup}(a). This AOM is the main source for noise, degrading our SNR$_\text{dB}$ measurements. In general, two sources of noise can be induced by an AOM: phase noise from the AOM driver and RF amplifier, and intensity noise. The intensity noise, which originates from relative diffraction efficiency fluctuations (RDEF) \cite{Liu_2018}, dominates in our setup. A measurement of the noise power on the spectrum analyser over a range of 50 to \SI{250}{kHz} with a resolution bandwidth of \SI{3}{\kHz} and for different probe beam powers is shown in Fig.~\ref{noisefloor}(a). The red data points represent a measurement in which the first order of the AOM ($\sim\SI{65}{\percent}$ efficiency) was detected on the photodetector (Thorlabs PDA 100A-EC, bandwith up to \SI{2.4}{\MHz}) and the optical power incident on the AOM was varied. We find that the noise power scales as quadratic function of the probe power. In a consecutive measurement (blue) the AOM was removed from the beam path, resulting in a linear increase of noise power with increasing probe power. A shot noise limited scenario (\SI{3}{dB} gain above the detector noise floor) is reached for probe powers $>\SI{4}{mW}$.

The intensity noise of the AOM limits the achievable SNR$_\text{dB}$ as depicted in Fig.~\ref{noisefloor}(b) (blue points). Towards higher probe powers the peak height of the AM signal, see Fig.~\ref{spectrumscan}(a), \rc{reaches its maximum} while the noisefloor increases quadratically with the probe power, resulting in a fall-off of the SNR$_\text{dB}$. If we compensate for the increasing noisefloor and only consider shot noise (black points) the SNR$_\text{dB}$ \rc{reaches a maximum of} $\sim\SI{32}{dB}$, and overall a higher SNR$_\text{dB}$  can be obtained. In order to reduce the intensity noise of the AOM the RF driving power of the AOM can be set close to its saturation point. To further reduce the noise, alternative options to generate multiple beams can be employed, as discussed in Sec.~\ref{Sec:3c2}. For a photon-shot-noise-limited measurement an optical heterodyne detection scheme can be employed, where a local oscillator beam is mixed with the transmitted probe as e.g. used in \cite{Meyer_2018}.

\section{BW and data capacity for multiple receivers} \label{AB}
\begin{figure}[t]
\centering
\includegraphics[width=0.98\columnwidth]{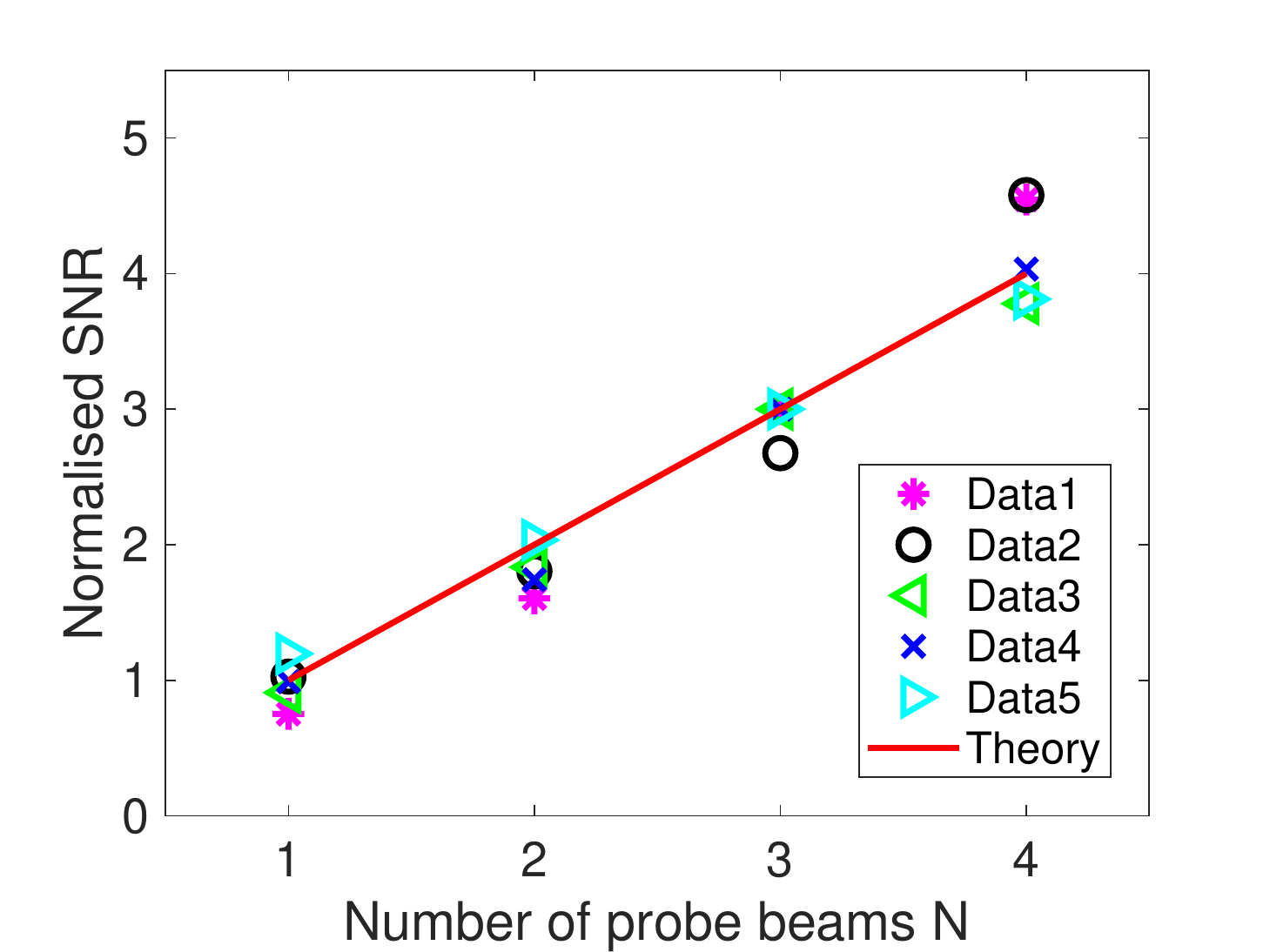}
\caption{\rc{Scaling of the SNR in means of a optical power ratio with the number of probe beams (receiver volumes). The data is normalised to the SNR of a single probe beam.  } \label{slopeSNR}}
\end{figure}

For a SIMO configuration with $N$ identical receiver volumes (i.e. $N$ probe beams) the signal-to-noise ratio in decibel is given by by
\begin{equation} \label{B1}
\text{SNR}_\text{dB} (N)= \text{SNR}_{\text{dB},N=1}+20 \times \text{log}_{10}(N).
\end{equation} \label{B2}
\rc{This implies that every time the number of receiver volumes, $N$, is doubled, the SNR in terms of a optical power ratio increases by a factor of two, as can be seen in Fig.~\ref{slopeSNR} for five sets of measurements. These measurements have in common that the SNR of the individual receivers were roughly identical, while the probe power of the beams could vary significantly from receiver to receiver.} 

The scaling in $\text{SNR}_\text{dB} (N)$ relates to an improvement of bandwidth with $N$ as
\begin{equation}
\text{BW}(N) = \text{BW}_{N=1} + 20/m \times \text{log}_{10}(N),
\end{equation}
where $m$ is the slope defined by the falloff of the SNR-curves towards higher AM frequencies in Figs.~\ref{sBeam}(a)-(b). The slope, $m$, depends on atomic parameters, such as coherence rates and vary for different experimental setups.

The achievable communication rate for a channel at a given amplitude modulation frequency $f_\text{AM}$ is given by the Shannon-Hartley theorem, Eq.~(\ref{SHT}), as $C = f_\text{AM}\times \text{log}_2(1 + \text{SNR})$. Equation~(\ref{B1}) can be used to derive the theoretical data capacity for $N$ receiver volumes
\begin{equation}
\begin{aligned}
C_A(N) &=f_\text{AM}\times \text{log}_2[1 + \text{SNR}(N)] \\
 &= f_\text{AM}\times \text{log}_2[1 + 10^{\text{SNR}_\text{dB}(N)/\rc{20}}]
 \\
 &\stackrel{\text{\ref{B1}}}{=}  f_\text{AM}\times \text{log}_2[1 + \rc{N} \times \text{SNR}_{N=1} ],
\end{aligned}
\end{equation}
\rc{where $\text{SNR}_{N=1}$ is the SNR of a single beam as ratio of optical power.}
\noindent

%

\end{document}